# Planning for Mission Efficiency, Robustness, and Resilience for Unmanned Autonomous Systems (UASs) in Contested Environments

Michael J. Gaiewski, Sergey N. Vecherin, Laurel P. Williams, Igor Linkov*

**Abstract** Unmanned Autonomous Systems (UASs) can remotely sense targets across large-scale areas but may be vulnerable to threats in contested environments. Current approaches consider efficiency through UAS path planning, but missions prioritizing robustness (i.e., the capacity to withstand disruptions) and resilience (i.e., the ability to recover from disruptions) are rarely considered despite their ability to increase mission success. This paper presents robust and resilient strategies for route planning for a fleet of UASs in a heterogeneous threat landscape. By integrating components of robustness and resilience in route planning through allowing the UASs to overlap in the waypoint locations and dynamic mission adaptation, the probability of mission success of complete area coverage can be increased. The results show that a robustness-focused strategy shows significant improvement in probability of mission success over an efficiency-based strategy, and that a resilience-based strategy outperforms both the efficiency- and robustness- based strategies. The specified methodologies provide promising future directions of research to explicitly incorporate robustness and resilience into UAS mission planning.

## I. INTRODUCTION

UNMANNED *Autonomous Systems* (UASs) are widely used to carry out a variety of tasks, including surveillance, reconnaissance, and intelligence gathering in risky or contested environments [1]. In these situations, UAS missions can transfer risk from human personnel to UASs while utilizing more efficient data collecting technologies [2]. In concert with increasing use of UAS, counter-UAS technologies have also rapidly developed [3]. There are therefore both incidental and malicious risks inherent to deploying UASs: malfunctioning, disconnection, or damage from complex weather and terrain conditions, or damaged or captured by an adversary. These risks could prevent UASs from collecting complete field data or expose sensitive information through data leakage.

To successfully conduct UAS operations in risky environments, special path planning techniques, such as coverage path planning [4], [5], need to be implemented to account for threats or dangers. These fall under the category of risk-based approaches, which aim to mitigate and prevent disruptions to collecting field data from occurring in the first place [6]. Path planning techniques that consider threats has been studied extensively [7], [8], [9]. In [10], a literature review was conducted on coverage path planning algorithms in static and dynamic risk environments, highlighting grid-based, graph-based, sampling-based, heuristic, adaptive, and real-time methods. In [11], UAS delivery missions prioritizing safer paths using various mathematical techniques were studied. In general, however, these methods are designed to optimize route or mission efficiency rather than explicitly prioritizing the success of data collection objectives under UAS attrition.

Prioritizing robustness- and resilience- based perspectives to UAS route planning in complex threat environments has only limited discussion in the literature [10]. *Robustness* is defined as the "capacity to withstand disruptions without damage to system functionality" [12]. A robust system can withstand disruption without the need for reconfiguration or intervention. *Resilience* is the ability of a system to "prepare and plan for, absorb, recover from, and more successfully adapt to adverse events" [13]. Such robust and/or resilient systems are better able to maintain their *critical function* ("a system function identified by stakeholders as an important dimension by which to assess system performance" [14])

This paper was submitted for review on January 15, 2026. This work was supported by the US Army Corps of Engineers.

M. J. Gaiewski is with Credere Associates, LLC., Westbrook, ME 04092, USA (email: Michael.J.Gaiewski@usace.army.mil).

S. N. Vecherin is with the Cold Regions Research and Engineering Laboratory, US Army Engineer Research and Development Center, Hanover, NH 03755, USA (email: Sergey.N.Vecherin@usace.army.mil).

L. P. Williams is with Credere Associates, LLC., Westbrook, ME 04092, USA (email: Laurel.P.Williams@usace.army.mil).

I. Linkov* is with the Environmental Laboratory, US Army Engineer Research and Development Center, Concord, MA 01742, USA (email: Igor.Linkov@usace.army.mil).

*Corresponding author

before, during, or after a disruption has occurred. Robustness- and resilience-based approaches to UAS route planning accept that disruption will occur and attempt to still succeed in the operation.

A use case or *mission* of UASs needing to survey an *entire* area may be important when critical information is scattered sparsely throughout an area of interest. In this case, the critical functionality is collection of the complete spatially distributed data to the home base without UAS damage or interception, either by mechanical failure or adversarial attack. A constraint added to the use case is that UASs cannot communicate with each other or with the base without physically being present at the base. Such a constraint is important when there is significant electronic warfare present in the area [15]. Such a constraint is also useful for missions where UASs must travel far distances that may go outside a practical range of a wired communication tether attached between a UAS and the base [16]. Under these constraints, UASs do not have the ability to change their initial objectives even in response to changing mission conditions, as doing so would require communication from another UAS or from the base. The mission robustness is determined by the success of information gathering under this approach, even in adverse conditions. But robustness will fail at sufficiently high levels of disruption; a resilient approach would allow for system reconfiguration to recover the mission once the disruption has occurred.

Previous work that implicitly assumes inter-UAS and/or base communication *during* the mission has been examined. For example, the authors in [17] and [18] each present resilient complete coverage path planning approaches by allowing for the reallocation of UASs if any in the fleet are disabled during the mission. Robust path planning methods that do not necessarily require these communications are also studied in the relevant literature. Some require that multiple UASs visit specific data collection locations at different times to ensure that if any in the fleet are disabled, the data is not lost [19]. However, the method implicitly assumes that once data is collected by a UAS, it is instantly transmitted to an operator and does not consider whether the UASs return to base unharmed. The study in [20] also employs robustness in that the authors' path planning strategy minimizes the loss of information in the case of a disruption to a UAS. However, this method does not guarantee complete area coverage, as is desired. These two references both define their approaches as *resilient* instead of *robust*, but based on the distinctions presented in this paper, these approaches fall more under the category of *robustness*.

Typically, optimization-based route planning algorithms are focused on determining the single best route (for example, the shortest, fastest, or safest one) [5]; however, deploying multiple UASs on this single route does not prioritize robustness or resilience. If all UASs travel in tight proximity to each other, and follow essentially the same route as a cluster, all UASs might be disabled simultaneously. Even if multiple UASs are sent one after another by the same best route, it still does not increase chances for success because an adversarial counter-UAS system will be tailored to the already known route. Instead, the UASs should be deployed on sufficiently different paths from each other to avoid this situation. This will mean, however, that the UASs cannot all take the same optimal path, and therefore suitable suboptimal paths must be determined. From a statistical perspective, sufficiently different routes make it reasonable to assume that the failures of individual UASs are mutually independent events. Disabling one UAS on its path does not affect chances to disable another UAS following a completely different path.

This paper will describe methods that rigorously combine optimality, robustness, resilience, and the probabilistic nature of unforeseen threats. This method uses mathematically rigorous and approximate methods to devise the optimal number of UASs and their routes to fulfill the task with a desired probability of success. Specifically, the following three main questions are addressed in this paper:

- How can one formulate the complete coverage surveillance objective into a mathematically rigorous computational model?
- What is the optimal path for each UAS deployed in the mission that minimizes the risk and yet is varied enough to avoid all UASs following the same path?
- What is the minimal predicted number of UASs needed to ensure a high likelihood of mission success that prioritizes robustness and/or resilience to the event of one or more UASs being lost or captured?

To answer these questions, several existing mathematical and risk-based path planning techniques will be synthesized together. A strategy focused on efficiency is presented first as a baseline comparison. Next, a robust focused strategy is presented which compares very favorably to the first strategy and specifically predicts the minimal number of UASs that must be deployed to ensure that the operation will meet a user-defined threshold probability of success even under unforeseen disruptions. Lastly, an enhancement is made to the efficiency-based strategy to develop a resilient-focused strategy, and it is shown to outperform both the previous two strategies. These results show promising directions for future research in developing more rigorous robust- and/or resilient-based mission planning strategies.

## II. Methodology

### A. Use Case

In this paper, a realistic use case of deploying a fleet of UASs from a home base to survey and collect imagery or other sensor data from all parts of a given area in the contested environment is studied. Such UASs can be aerial-, ground-, or aquatic-based, among other types. For example, aerial-based UASs may be sent to determine the locations of mines in a minefield [21]. They may also be sent to survey a disaster area for the search and rescue of survivors [22], or even need to take action in the field like fighting a wildfire (see Section IV-B below). In such applications, it is important that all parts of the area are surveyed without blind spots.

This exercise assumes that an accurate two-dimensional grid *map* of the desired area is available to provide an associated probability of a UAS failure at each map pixel, e.g., due to passive hazards or active interference. The altitude of the UAS is not considered but could be incorporated into such risk probabilities. Literature on methodologies for approximating these exact maps in contested areas is limited, however there are several numerical approaches that serve as reasonable starting points. In [23], for example, fundamental research on determining grid-based probability maps mainly based on environmental hazard is performed. For the considered examples, a synthetically generated map is used (Fig. 1), where intensity of the grayscale colormap represents the probability of failure; the darker the pixel the higher probability that a UAS will be disabled if passing through it. Subsequently, the complementary success probabilities are obtained as the difference between one and the probability of failure. In Fig. 1, it is worth noting that the minimum probability of failure on each pixel is 0.001.

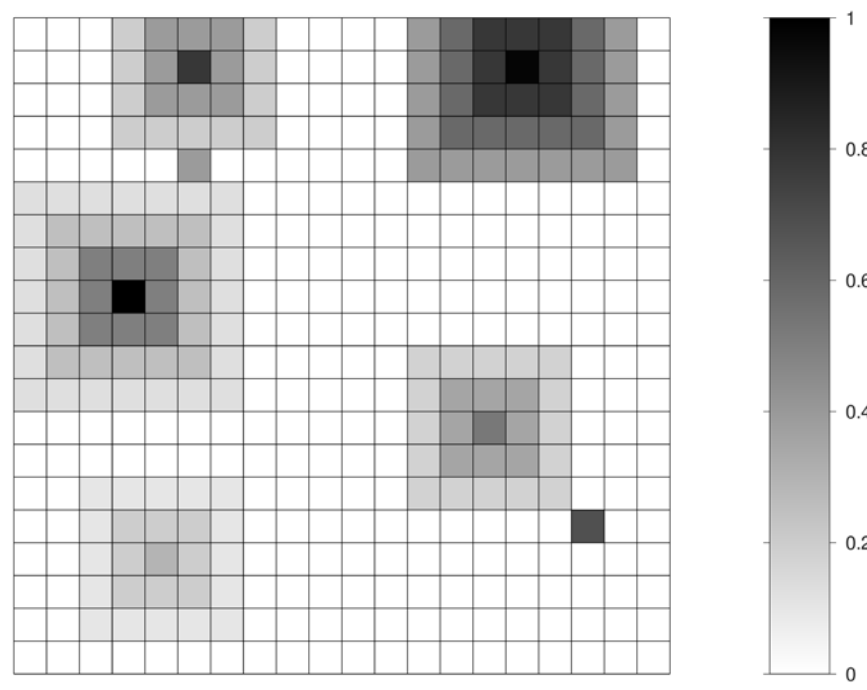

Fig. 1. Example of a map with risk probability of traversal associated to each pixel.

The UASs will be preprogrammed to travel autonomously on different routes that start and end at the home base, which is a key requirement for the considered mission. Each UAS can be assigned to part of or the entire map to surveil, as shown in Section II-B. It is assumed that each UAS cannot communicate with any other and that the only way each UAS can communicate with the base is by physically returning to the base. Additionally, the sensors carried by the UASs have known ranges of sensing around the UAS and can view any pixel on the grid that falls in that radius.

For example, in Fig. 2, if a UAS is located at the pixel with the blue dot, it can view every pixel on the map with a red dot on it. (For the purposes of these calculations, sensor range here is assumed to be circular, with a radius of 5 pixels. However, a different shape or radius can be defined depending on the UAS in question.) If there are obstacles that obstruct a UAS's view, they will alter the sensing footprint of the UASs by making it irregular, but this nuance does not principally limit the presented framework and, therefore, obstacles are not considered in the current study. For example, if an obstacle is known, the UAS surveillance radius could be blocked at certain pixels.

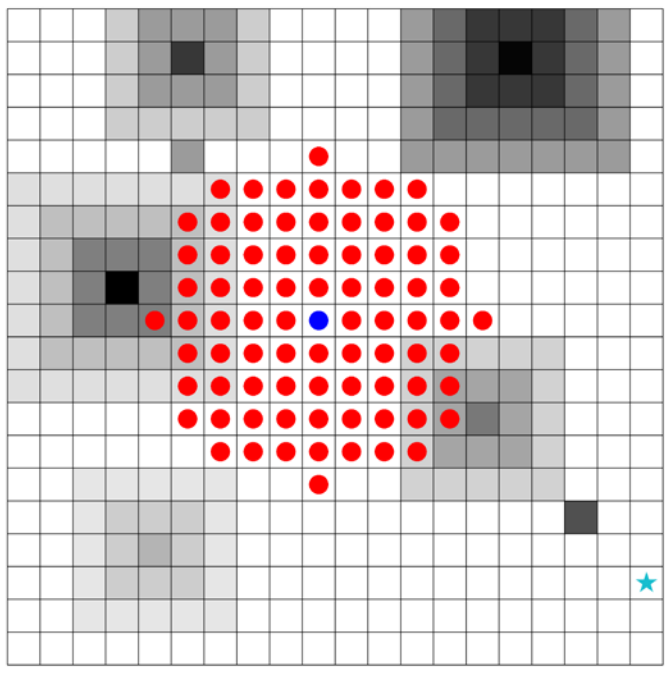

Fig. 2. Example of the sensing range of a UAS. If a UAS is located at the pixel with the blue dot, red dots depict its sensing range. The base is located at the cyan star.

The mission is considered successful if information from *every* pixel on the map is delivered to base, where the information is then integrated and processed. Note that it does not imply that every UAS needs to return to the base. For these experiments, the base is placed on the pixel in the 18th row and 20th column (the cyan star), and its threat probability is set to zero (Fig. 2).

Three path planning algorithms in Section II-B are developed with various mathematical techniques and comparative analysis of these two strategies will be given. The first strategy consists of dividing the UASs into different groups assigned to survey different sections of the map, and the second is that all UASs visit all parts of the map but on intentionally different paths and different times. The third strategy is a natural enhancement on the first strategy that allows for surviving UAS redeployment. All three strategies share some common techniques which are presented in the remainder of Section II-A.

*1) Probability of Successfully Traversing a Path*

Since each pixel on the grid has an associated threat probability, then any path has an associated risk of traversing through it. If a UAS crosses $T$ pixels along its path, the probabilities of *success* on each pixel are $p_1, \ldots, p_T$, and the events of successfully traveling through each pixel are probabilistically mutually independent, then one can multiply the probabilities of success together to get the probability $P_T$ that a UAS will successfully traverse a path or route:

$$P_T = \prod_{j=1}^{T} p_j. \quad (1)$$

*2) Waypoints*

One approach to efficiently collect data from all parts of the map is to determine the minimal number of *waypoint* locations such that if all waypoints are visited and data is collected only from each waypoint location, then every pixel on the map will be surveilled. These waypoints should also be placed in locations that are "safe" to visit within a reasonable threat tolerance. It is worth noting that setting too strict of a threat tolerance may make the problem infeasible: in the upcoming numerical experiments, the threat tolerance was set to 1%, though this can be changed based on user preferences. Also note that the waypoint locations are determined independently of wherever the base is located. The objective is to find the minimum number $W$ of waypoints and their optimal locations that satisfy these desired characteristics. This is a classical computational problem known as the *Set Cover Problem* [24]. It can be solved via integer linear programming. A full description of the mathematics of this problem is described in Appendix Section A-1). This problem is known to be *NP*-hard, so computational efficiency scales rather poorly with the size of the map. More computationally efficient approximation methods are discussed in Appendix Section A-2), though they were not necessary in the example problems considered in this paper to achieve a reasonable computational time compared to the exact algorithm.

#### *3) Best Paths Between Waypoints*

An algorithm is needed to find the safest paths between every pair of waypoints, including the base. The threat maps that have been used are functionally similar to *occupancy maps* when it comes to route planning [25]. Each pixel in this case represents the probability of failure in crossing that pixel; similarly, each pixel on an occupancy map represents the probability an obstacle is present at that pixel. The *A* algorithm* as described in [26] can be utilized to plan routes accounting for these risks. A MATLAB/Octave implementation of the *A** algorithm developed in [27] was modified to avoid dangerous areas, unless it is completely necessary to travel through one. Full details on how the *A** algorithm was modified are presented in Appendix Section B.

### *B. Strategy Comparison*

Determining the best way to have the UASs visit the waypoints and return to base can be implemented using multiple strategies. Three such strategies are presented in the following three subsections.

#### *1) Efficient Strategy*

One strategy to solve the problem considered is to split the fleet of $N$ UASs up so that each UAS is assigned to visit a specific group of waypoints in a specific order and return to base. Ultimately, with enough UASs available, each UAS can visit just a single waypoint, collect data at that location, and bring it back to the base. This strategy is motivated by trying to minimize the overall mission time to successfully gain information as quickly as possible, which is used in practice [28]. From this point forward, this strategy is referred to as the *Efficient* strategy.

The success probabilities between each pair of waypoints, including the paths from the base to each waypoint, can be used as the cost function to set up an implementation. A modification of the *Multiple Traveling Salesperson Problem* (MTSP), its original form described in [29], was used. This is a famous problem that can be solved using integer linear programming. The MTSP typically seeks to find the shortest paths; however, the problem considered in this paper will require a "maxmin" objective of maximizing the lowest individual UAS probability of returning [30], [31], [32]. Additionally, a multi-objective component was added to consider both maximizing the probability of the lowest UAS returning and balancing the distances each UAS travels. This is a realistic strategy in practice because the mission may need to be done on a time constraint, so maximizing probability of return and minimizing the overall mission time is desirable [28].

Since the MTSP, as well as this maxmin and multi-objective version of the problem, are *NP*-hard problems, approximation algorithms are often needed for computational efficiency on large maps, though they were not necessary for this given example. A technical description of this implementation of the multi-objective MTSP is contained in Appendix Section C, and an example of an approximation algorithm is described in Appendix Section E.

In this strategy, the success of the overall mission hinges on whether *all* UASs that are deployed return to base, because no proper subset of UASs possess the information about the entire map. It is assumed that the success or failures of each UAS are mutually independent because this strategy separates the UASs into completely different sets of waypoints, so there is little to no overlap in their paths in space or in time. Under this assumption, the probability of a successful mission is simply the product of the probabilities of each of the deployed UASs returning.

#### *2) Robust Strategy*

The so-called *Robust* strategy deploys a fleet of $N$ UASs in such a way that *all* UASs visit *all* waypoints but using sufficiently different routes, which implies different order and time of visiting each waypoint. This maintains the probabilistic independence of UAS failure and avoids simultaneous disablement as a group. This independence is assumed since UASs will, in general, not travel along the same paths at the same *times* or in the same *directions*. Even though the paths between waypoints are always the same, the forthcoming *variety constraint* largely forces the complete paths to differ in time and the order of waypoint visits.

This strategy exemplifies the robust approach to the mission. It combines several quantifiably different imperfect routes to achieve higher probability of mission success than any single route individually. To implement the Robust strategy, variations to the original *Traveling Salesperson Problem* (TSP) as seen in [32] are considered: here, the maximization of the probability of at least one UAS to return is considered. This is another well-known *NP*-hard problem that can be solved with integer linear programming, and where approximation methods might be used if necessary.

The original TSP formulation can only find the best singular route. However, the goal is to find $B$ best different enough routes to choose from and then determine the minimal number ($N \leq B$) of these to assign to UASs to ensure a desired likelihood of mission success. To accomplish this, the algorithm is run $B$ times, adding a constraint to the problem at each iteration that disallows routes that are quantifiably similar to the previously found routes. For this experiment, $B$ was set to 7, though it may be set differently depending on user preferences. However, if $B$ is too low, a given desired probability of a successful mission may not be obtainable. More technical details on this method can be found in Appendix Section D, with approximation methods being discussed in Appendix Section E.

A mission is considered successful using the Robust strategy if *one* or more UASs return to base since each UAS possesses information about the entire map. Here, robustness is achieved largely through the redundancy of each UAS route. The mission is designed to succeed even if a disruption of up to $N$-1 UASs being disabled occurs. The rest of the fleet will still be scheduled to visit all the waypoints, which is expected to increase the probability of a successful mission. The algorithm can determine the minimal $N$ value so that at least one UAS will successfully return to base with the probability greater than a desired threshold, $x$%, where $x$% is defined by the user. The calculations are described in detail in Appendix Section F. For the described numerical experiments, $x = 99\%$.

*3) Resilient Strategy*

A natural approach to hybridize the two strategies (referred to as the *Resilient* strategy) would be to start with the Efficient strategy and for any UASs that do not return to base, redeploy surviving UASs on paths whose information was not successfully collected until either all UASs are disabled, or all information is collected. Resilience in this strategy comes from reconfiguring the mission plan to recover the critical function of achieving full map coverage. The mission can still be successful even after the disruption has occurred. Such a strategy, however, may require in practice a high preparation time, which may not be feasible in mission-critical time settings. Surviving UASs would need to be prepared again for redeployment, which involves recharging, upkeeping, and route reprogramming. Additionally, as mentioned in the Introduction, an adversary may notice UASs being deployed one after another along the same routes and take additional measures to disable the predictable upcoming UASs on those routes. However, if preparation time is negligible, and the threat probability of the same revisited routes does not increase, numerical simulations are performed to serve as a proof of concept to a hybrid strategy. For $N = 1,\ldots,7$, 10,000 Monte Carlo computational mission simulations were run using the theoretical routes and probabilities that the Efficient strategy estimated. Each simulation treated the traversal of each pixel on a path for each vehicle as a Bernoulli random variable. That is, for each pixel traveled, a random number between 0 and 1 was generated and if it was above the threat probability, the vehicle successfully traversed that pixel, and the vehicle failed on that pixel otherwise. For any surviving UAS that returns to base, it is assigned a random route for a UAS that did not survive. It is also assumed that if any two or more UASs are assigned to the same route, they are not deployed at the same time to maintain probabilistic independence.

## III. RESULTS

### *A. Waypoints*

Using the same threat map as seen in Fig. 1, and the same coverage shape and radius as seen in Fig. 2, the results for the optimal waypoint locations are presented in Fig. 3. The exact integer linear programming algorithm found that the minimal number of waypoints $W$ needed was 9.

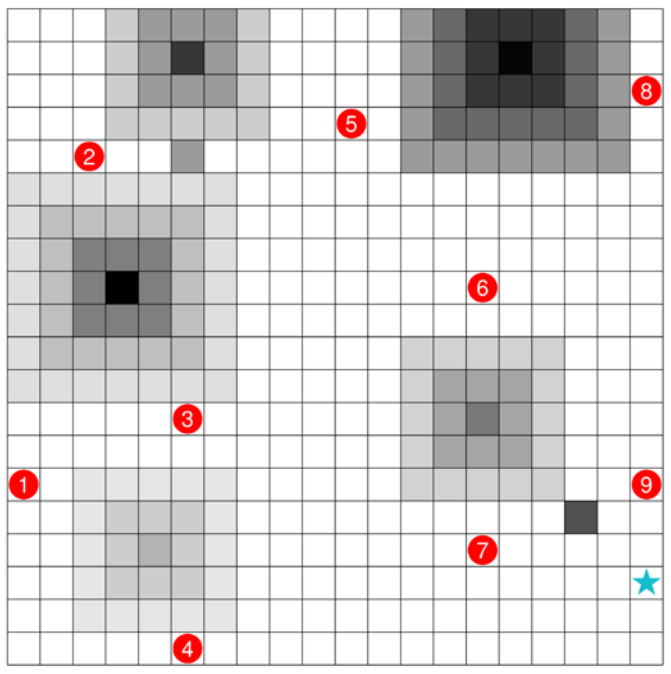

Fig. 3. Optimal waypoint locations from integer linear programming given the threat map and coverage shape of Figs. 1 and 2.

If the problem is indeed feasible, the exact algorithm will always find a solution that satisfies the desired conditions. Note that the solution is not unique: other linear programming solvers may find different sets of waypoint locations with the same value of objective function ($W = 9$ for this example), while approximate algorithms may find solutions with a higher value.

### *B. Best Paths Between Waypoints*

The optimal paths between each pair of waypoints were determined using the modified $A^*$ approach described in Subsection II-A-3) and Appendix Section B. The probabilities of successfully traveling from any two pairs of waypoints are shown in Table I. Diagonal elements in Table I indicating "traveling" from a waypoint to itself are removed.

Table I: Probability of successfully traveling from each pair of waypoints.

| | Base | 1 | 2 | 3 | 4 | 5 | 6 | 7 | 8 | 9 |
|---|---|---|---|---|---|---|---|---|---|---|
| Base | -- | 0.9822 | 0.8572 | 0.9871 | 0.9871 | 0.9871 | 0.9910 | 0.9960 | 0.9861 | 0.9980 |
| 1 | 0.9822 | -- | 0.8606 | 0.9960 | 0.9930 | 0.9861 | 0.9871 | 0.9871 | 0.9802 | 0.9822 |
| 2 | 0.8572 | 0.8606 | -- | 0.8649 | 0.8589 | 0.8693 | 0.8658 | 0.8615 | 0.8598 | 0.8598 |
| 3 | 0.9871 | 0.9960 | 0.8649 | -- | 0.9930 | 0.9910 | 0.9920 | 0.9920 | 0.9851 | 0.9871 |
| 4 | 0.9871 | 0.9930 | 0.8589 | 0.9930 | -- | 0.9841 | 0.9881 | 0.9920 | 0.9812 | 0.9871 |
| 5 | 0.9871 | 0.9861 | 0.8693 | 0.9910 | 0.9841 | -- | 0.9960 | 0.9871 | 0.9900 | 0.9900 |
| 6 | 0.9910 | 0.9871 | 0.8658 | 0.9920 | 0.9881 | 0.9960 | -- | 0.9910 | 0.9940 | 0.9940 |
| 7 | 0.9960 | 0.9871 | 0.8615 | 0.9920 | 0.9920 | 0.9871 | 0.9910 | -- | 0.9861 | 0.9960 |
| 8 | 0.9861 | 0.9802 | 0.8598 | 0.9851 | 0.9812 | 0.9900 | 0.9940 | 0.9861 | -- | 0.9891 |
| 9 | 0.9980 | 0.9822 | 0.8598 | 0.9871 | 0.9871 | 0.9900 | 0.9940 | 0.9960 | 0.9891 | -- |

These are only the *probabilities* of success of traveling strictly between any pair of waypoints. There is still an associated success probability *visiting* each waypoint, so the waypoint's success probability needs to be additionally multiplied in when computing the probability of a successful path.

### C. Efficient Strategy

For the Efficient strategy, the best sets of routes from deploying $N$ UASs were determined for $N$ = 1,…,7. For legibility, the best routes for $N$ = 3 are plotted in Fig. 4.

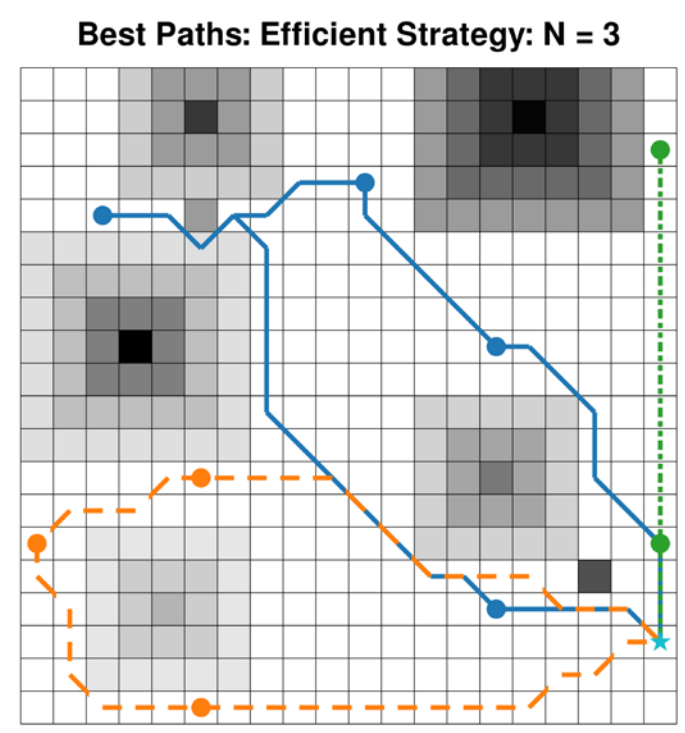

Fig. 4. The best $N$ = 3 routes using the Efficient strategy. The cyan star represents the base, and the blue/solid, orange/dashed, and green/dotted waypoints and paths represent the routes that the 3 UASs traveled on.

As $N$ increases and reaches the number of waypoints $W$, the algorithm will tend towards setting each route to each individual waypoint. Furthermore, the algorithm will not be feasible to run if $N$ exceeds the number of waypoints $W$. Like the waypoint results, each solution as $N$ iterates from 1 to 7 is not necessarily unique. Like the waypoint problem in Section III-A, other linear programming solvers may find different solutions but will not find solutions with a higher probability of success than these ones.

The strategy requires all $N$ deployed UASs to return to base to achieve mission success. Assuming their statistical independence, the probabilities of each UAS returning are multiplied to find the probability of success of the entire mission. The probabilities were 0.7333 (blue/solid), 0.9608 (orange/dashed), and 0.9714 (green/dotted), resulting in a successful mission probability of 0.6844.

### D. Robust Strategy

In the Robust strategy, the probabilities of each of the first $B$ = 7 routes are shown in Table II. Note that these 7 best routes have individual success probabilities ranging from 0.7003 to 0.7116, less than all three return probabilities in the Efficient strategy.

Table II: Return probabilities of each route returning using the Robust strategy.

| Route # | Probability of Route Success |
|---|---|
| 1 | 0.7116 |
| 2 | 0.7116 |
| 3 | 0.7059 |
| 4 | 0.7059 |
| 5 | 0.7031 |
| 6 | 0.7031 |
| 7 | 0.7003 |

The top 4 routes corresponding to Table II are shown in Fig. 5. Note all routes start and end at the base; the first waypoint is marked by the red dot, and the last visited waypoint is blue, following a color gradient from red to blue and numerical sequence along the way. One immediate observation is that Paths 2 and 4 are the reverse of Paths 1 and 3, respectively.

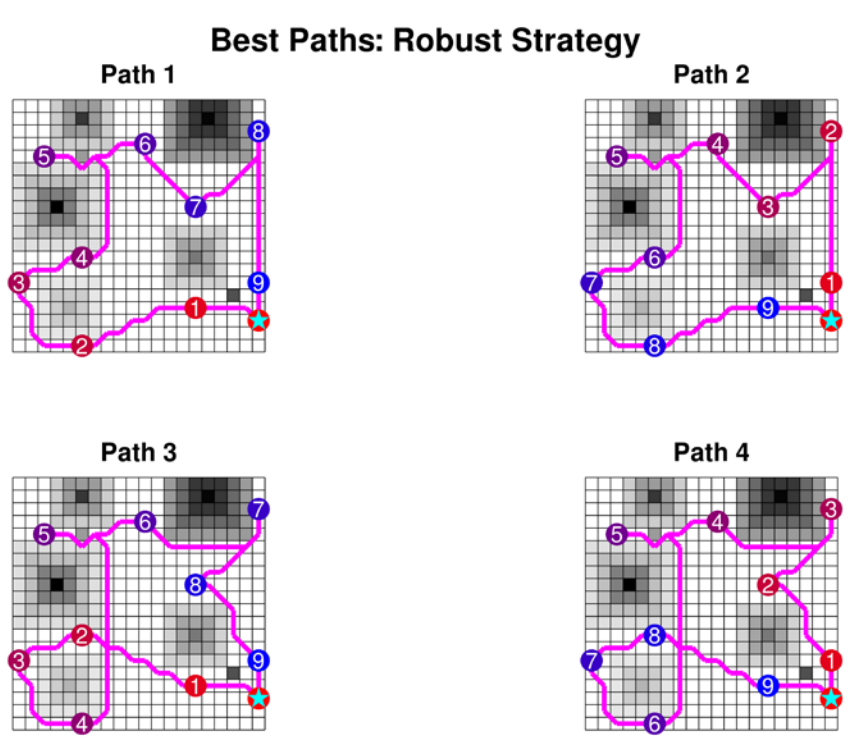

Fig. 5. Top four UAS routes to take using the Robust strategy. Routes start and end at the base (cyan star). The route direction follows a color gradient from red to blue and numerical sequence along the way.

The mission is successful if at least one UAS returns to base, as opposed to the Efficient strategy, which requires all UASs to return to base. The exact formula for determining the probability of a successful mission under this strategy is described in Appendix Section F. For this example, for $N$ = 3, the probability of a successful mission is 0.9755. It is interesting to note that the Robust strategy, operating with routes that are all riskier than the three routes in the Efficient strategy, produces a result with significantly higher probability of success. This is a nontrivial and nonobvious outcome that is achieved using the robustness concept in route planning. Again, the non-uniqueness could affect the recommended paths and their probabilities of success on later iterations as the route variety constraint may forbid different paths depending on the specific linear programming solver that is used.

### E. Resilient Strategy

The probabilities of a successful mission from the 10,000 Monte Carlo Resilience strategy simulations are shown in Table III.

Table III: Simulated probabilities of a successful mission across Monte Carlo 10,000 simulations using the Resilient strategy.

| # UASs ($N$) | Probability of Mission Success |
|---|---|
| 1 | 0.7188 |
| 2 | 0.9128 |
| 3 | 0.9795 |
| 4 | 0.9934 |
| 5 | 0.9981 |

| 6 | 0.9995 |
|---|---|
| 7 | 0.9997 |

The results show that the probability of a successful mission increases monotonically to 1 as $N$ increases (consider Appendix Section F, letting all $p_i$ be identical for *each* of the routes). Like with the Robust strategy, the user can choose a minimal $N$ that achieves a desired probability of mission success. ($N$ = 4 in this example). Because this strategy involves stochastic modeling and random redeployment assignments over multiple redeployment cycles, deriving an exact analytical expression for the probability of mission success is beyond the scope of this paper.

### *F. Comparison Between the Three Strategies*

#### *1) Theoretical Comparison (Efficient and Robust)*

Given the probabilities of a successful return using the Robust strategy shown in Table II, and letting $x$ = 99%, the strategy found that $N$ = 4 is the minimum number of UASs to deploy to ensure a successful mission greater than 99%. Fig. 6 presents the probability of a successful mission using both strategies with $N$ UASs ranging from $N$ = 1,…,7.

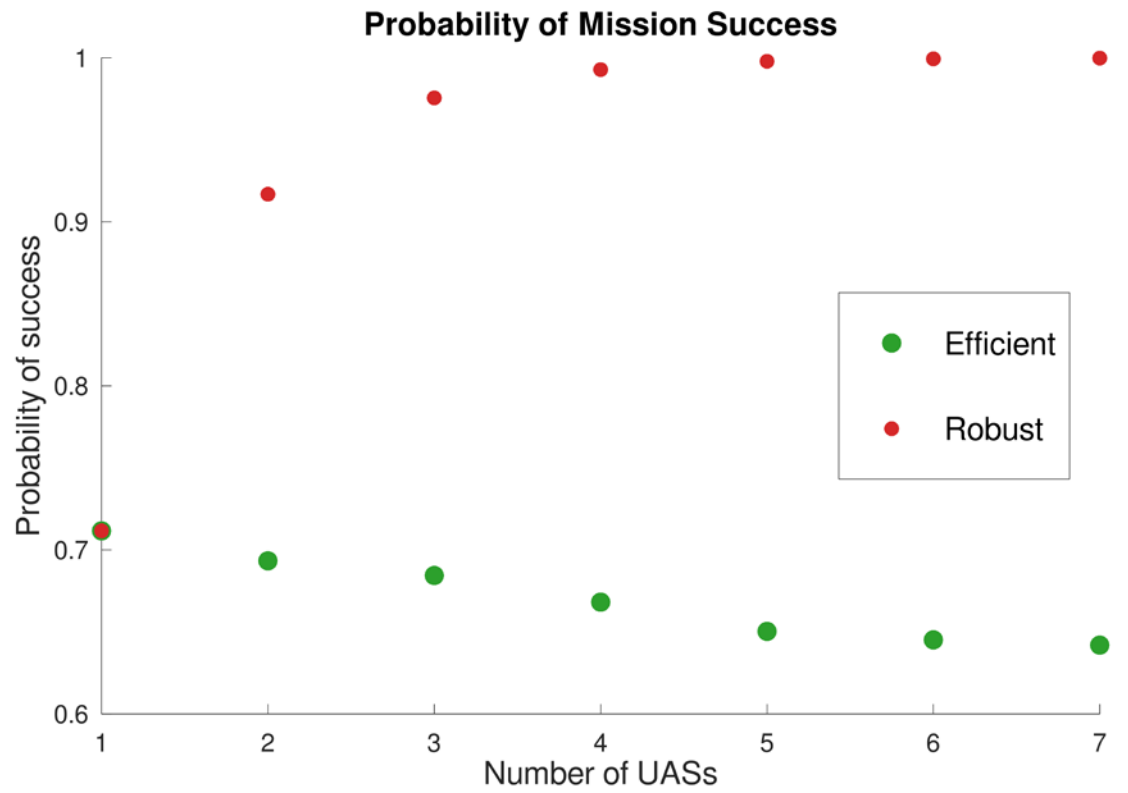

Fig. 6. Probabilities of mission success of the Efficient and the Robust strategies.

One observation is that at $N$ = 1, both strategies end up using the same best route for a single UAS, so they have the same probability of success. However, using the Robust strategy, the probability of a successful mission increases as more UASs are added. In contrast, for the Efficient strategy, the probability of a successful mission decreases as more UASs routes are deployed. For some specific threat zone configuration on the map, there might be situations when the Efficient strategy will have the same probability of mission success, but this does not apply in general.

#### *2) Numerical Simulation Comparison*

Ten thousand of the same aforementioned Monte Carlo simulations were run to validate theoretical results for the Efficient and Robust strategies, as well as compare the results to the Resilient strategy. Table IV shows the theoretical probabilities of mission success for the Efficient and Robust strategies, as well as the numerically simulated probabilities of mission success for all three strategies.

Table IV: Theoretical and Monte Carlo simulated mission success probabilities for each strategy.

| #UASs | Efficient: Theoretical | Efficient: Simulation | | Robust: Theoretical | Robust: Simulation | | Resilient: Simulation |
|---|---|---|---|---|---|---|---|
| 1 | 0.7116 | 0.7188 | | 0.7116 | 0.7082 | | 0.7188 |
| 2 | 0.6933 | 0.6902 | | 0.9168 | 0.9191 | | 0.9128 |
| 3 | 0.6844 | 0.6897 | | 0.9755 | 0.9731 | | 0.9795 |
| 4 | 0.6681 | 0.6666 | | 0.9928 | 0.9931 | | 0.9934 |
| 5 | 0.6503 | 0.6475 | | 0.9979 | 0.9976 | | 0.9981 |
| 6 | 0.6452 | 0.6481 | | 0.9994 | 0.9993 | | 0.9995 |
| 7 | 0.6419 | 0.6426 | | 0.9998 | 0.9997 | | 0.9997 |

The simulated probabilities were consistent to the theoretical probabilities for the Efficient and Robust strategies. The probabilities of success between the Robust and Resilient strategies were very similar to each other and remarkably higher than for the Efficient strategy.

### *G. Simulated Cost of UAS Attrition*

The average number of UASs lost across the 10,000 simulations of the Robust and Resilient strategies are compared in Fig. 7.

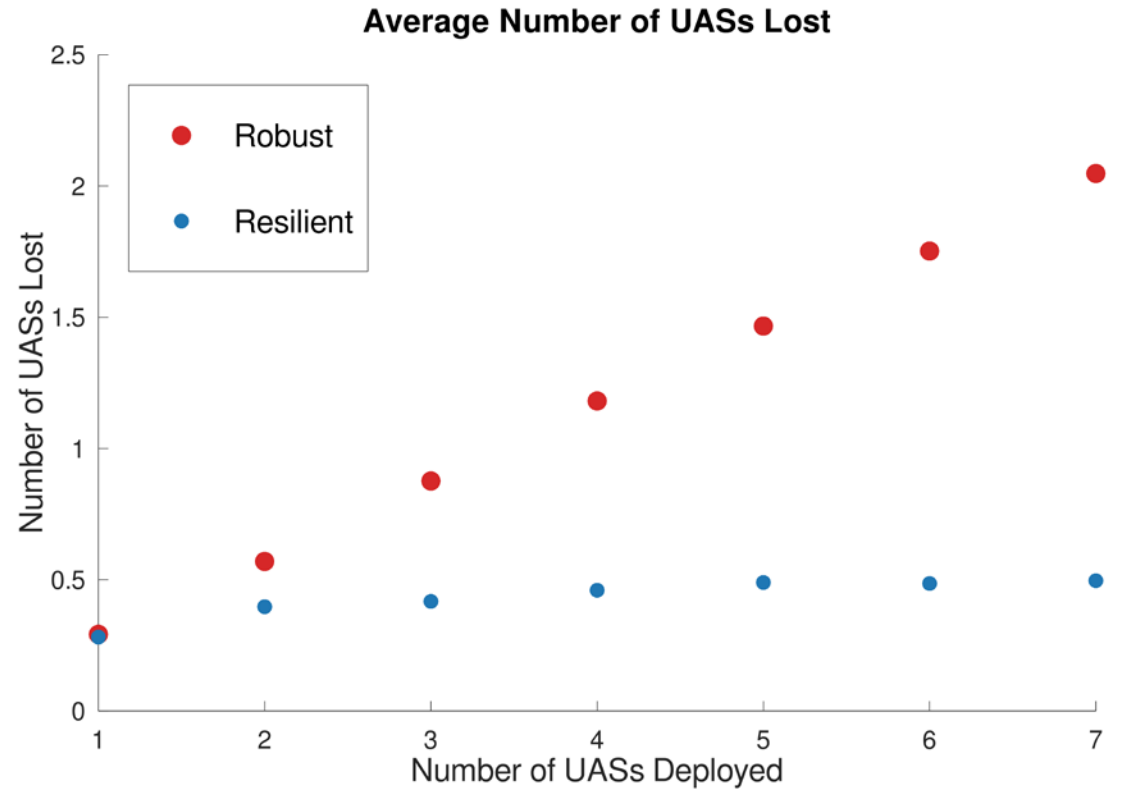

Fig. 7. Average number of UASs lost across 10,000 simulations of the Robust and Resilient strategies.

The results show that Resilient strategy outperforms the Robust strategy in terms of average number of UASs lost during a successful mission as the number of UASs increases.

### *H. Map Sensitivity Study*

It is assumed that the threat map provides an accurate estimate of the threat probabilities at each pixel. To show the sensitivity of the methodology to the input map, 10 randomly perturbed threat maps were created from the original threat map as seen in Fig. 1. A random probability value from [-0.02,0.02] was added to each pixel, except for the base, on the original threat map. Any perturbed probabilities that were below 0.001 were set to 0.001. The Robust strategy was performed on each of these 10 maps, with $B$ = 1,…,10 for each map. Table V shows for each perturbed map the number of waypoints needed, the minimum number of UASs ($N$) needed to achieve a mission probability of success of at least 0.99, and the theoretical probability of mission success at that $N$ value. Table V also shows the simulated probabilities of mission success and average number of UASs lost over 1,000 numerical simulations using the corresponding $N$ for the Robust and Resilient strategies.

Table V: Results from the 10 random map perturbations. Each row indicates the a distorted map version, the number of waypoints needed, the minimum number of UASs needed for the Robust strategy to achieve a theoretical probability of success of at least 0.99 ($N$), the obtained theoretical probability of mission success of the Robust strategy, and the simulated probabilities and average number of UASs lost for the Robust and Resilient strategies at this $N$ for 1,000 numerical simulations.

| Map | $W$ | $N$ | Prob. Success: Robust | Sim. Prob. Success: Robust | Avg. UASs Lost: Robust | Sim. Prob. Success: Resilient | Avg. UASs Lost: Resilient |
|---|---|---|---|---|---|---|---|
| 1 | 9 | 6 | 0.994 | 0.996 | 2.504 | 1.000 | 0.770 |
| 2 | 9 | 6 | 0.996 | 0.997 | 2.360 | 0.996 | 0.798 |
| 3 | 9 | 6 | 0.992 | 0.994 | 2.700 | 0.996 | 0.817 |
| 4 | 9 | 5 | 0.992 | 0.994 | 1.896 | 0.993 | 0.668 |
| 5 | 9 | 5 | 0.992 | 0.994 | 1.833 | 0.997 | 0.709 |
| 6 | 9 | 5 | 0.991 | 0.991 | 1.925 | 0.996 | 0.722 |
| 7 | 9 | 6 | 0.995 | 0.993 | 2.443 | 0.998 | 0.748 |
| 8 | 9 | 5 | 0.990 | 0.990 | 1.996 | 0.995 | 0.712 |
| 9 | 9 | 5 | 0.991 | 0.986 | 1.923 | 0.996 | 0.735 |
| 10 | 9 | 5 | 0.991 | 0.992 | 1.946 | 0.995 | 0.678 |

For this sensitivity study, the number of waypoints needed was not increased for any of the maps. Additionally, from this table, it is evident that perturbations in the input map can increase the number of UASs ($N$) needed for a successful mission with the Robust strategy. The Resilient strategy remains comparable to the Robust strategy in simulated probability of mission success and favorable in average number of UASs lost. This study indeed provides evidence that an accurate threat map is an important input parameter.

## IV. Discussion

### A. Results Interpretation and Connection to Risk, Robustness, and Resilience

Comparing the path lengths in Figs. 4 and 5, one can observe that each UAS individually travels a shorter distance under the Efficient strategy compared to the distances they travel in the Robust strategy. However, this does not make the Efficient strategy safer or as robust. A striking improvement is seen in return on resources invested by the Robust strategy versus the Efficient strategy: in the case of the former, the probability of a successful mission increases as more UASs are deployed, whereas in the case of the latter, the probability of a successful mission decreases. This difference occurs due to the definition of mission success as well as the application of robustness. In contrast, the Robust strategy deploys multiple UASs to substantially distinct routes and requires only one UAS to return to the base. The Efficient strategy can be enhanced to the Resilient strategy, and it performs overall favorably to the Robust strategy, particularly in terms of average number of UASs lost. Recalling that resilience is the ability to plan for, absorb, adapt to and recover from a disruption, the Resilient strategy allows for mission recovery even once the disruption has occurred, and still allows for a high probability of mission success.

Another important aspect of this mission design that mitigates the risk of the mission and supports resilience is the insistence on making the UASs travel on *sufficiently different* paths, (defined quantitatively in the Appendix Section D). The Efficient strategy inherently routes UASs on very different paths from one another, while the Robust strategy via its variety constraint enforces its different paths. This is a desirable and highly useful feature, because certain adversarial attacks [3] or unfavorable weather conditions could disable the entire UAS cluster at once if they were traveling together. From a statistical perspective, separating the UASs into sufficiently different paths approximately makes the individual failure of one UAS probabilistically independent from another.

Decision-makers may find this research useful because they can ensure that the mission will be successful to within $>x\%$ probability with the Robust strategy even with the disruption that one or more of the deployed UASs may go offline. These results suggest that intuitive or most efficient strategies (like following the same most optimal route) may not be the most robust or resilient. Robustness and resilience should be carefully incorporated into the mission design stage and quantitatively analyzed where possible.

### B. Use Case Variation: Fire Suppression

The use case considered does not simply have to involve data collection and surveillance. For example, the UASs may need to suppress a wildfire, at each waypoint deploying a fire suppression system payload that will affect the same surveillance radius [33]. Some UASs may be lost to the wildfire, so by deploying the minimal number of UASs such that at least one returns to base with $>x\%$ probability, the entire map area can be covered by the fire suppression system. Similar tasks arise with the lost personnel rescue missions or wildlife monitoring.

### C. Model Limitations

There are some limitations to the proposed methodology. In this initial effort, several assumptions are made that may not be realistic. First, it is assumed that a detailed map of threat areas and quantified risks of passing through this map are available. As mentioned above, fundamental research on these maps was reported in [23]; however, especially in dynamic situations, these estimates will involve uncertainty. A sensitivity analysis showed that small perturbations in the probability of each map can increase the number of UASs needed for a successful mission. Second, it is assumed that each UAS can view any pixel on the map within its surveillance range. In application, topography or human-built structures may block the view of certain map pixels depending on the UAS's location. While such obstacles may interfere with surveillance or routing, they do not diminish the proposed methodology in principle, as there are effective ways to account for obstacles in the proposed mathematical framework (i.e., limiting which pixels are visible depending on where a UAS is located). Another assumption is that UASs have enough physical power to travel the required distances. The map may have a large scale, and it may not be physically possible to have the UASs travel on the paths devised by either algorithm. In this case, each UAS's total allowed path length should be explicitly incorporated into the optimization as an additional constraint. Similarly, it is also assumed that UASs can make sharp turns. Typically, UASs travel in smoother arcs. However, on-board navigation systems of UASs are capable of handling sharp route corners and smoothing them out automatically.

As mentioned, using different linear programming solvers/packages may produce different routing results. Specifically, when determining the $N$ best paths, different paths across different linear programming solvers may disallow certain paths from being considered due to the stricter variety constraints with each iteration. More research is needed to study solutions across different solvers.

Finally, both strategies rely on the ability to solve *NP*-hard problems, which do not scale well in terms of computational efficiency. The approximation methods described in the Appendix provide one solution to this problem, but more research is needed to investigate and develop other approximation methods.

### *D. Future Work*

Future research could advance this effort in several areas. As previously mentioned, more realistic assumptions could be incorporated into the use case, and more approximate methods to solving the linear programming problems could be investigated.

Future work may also involve testing the strategies on entirely different input threat maps. Additionally, the work could be enhanced to incorporate techniques supporting a map having dynamically changing probabilities of risk. One limitation of the strategies explored in this research is that each UAS always takes the same path between any pair of waypoints. An adversary could observe this and react by intercepting UASs or moving hazards. In the case of a natural hazard such as fire or wind, hazards may not remain static. In both cases, initially selected paths may not remain the safest ones. Developing more dynamic techniques to respond to changing situations in the field could allow new safer paths to be determined during the mission.

Note that the Efficient and Robust strategies are essentially two ends of a spectrum, one where all waypoints are visited by all UASs and one where each waypoint is visited by only one UAS. More mathematically rigorous, as well as robust and/or resilient methods could be formulated that achieve a high efficiency mission, a high probability of mission success, and a low number of average UASs lost. The main advantage of the Robust strategy over the Efficient strategy is that waypoints are visited by more than one UAS, increasing probability of mission success when UASs are lost. If a method were developed to allow for multiple UASs to visit only higher-risk waypoints, this could blend the faster mission runtime of the Efficient strategy with the high probability of mission success of the Robust strategy. Some waypoints, for example, may be safe enough where only one or two UASs need to visit them and immediately return to base instead of requiring all UASs to visit each of them.

The preliminary Resilient strategy presented may also be improved upon. For one, it simply redeploys UASs on the exact same paths from where UASs did not return to base. A more mathematically rigorous redeployment strategy may be studied that better divides up remaining waypoints. This would likely avoid increased risk of an adversary enhancing hazards on paths that they deduce future UASs will be traveling. The presented favorable results of this strategy show promising potential for this future direction of research and show that resilience-based approaches should also be considered in mission planning.

Additionally, methods could be explored to divide the map into sectors that each would be surveyed by a subset of multiple UAS. In both cases, developing additional methods can enhance understanding of how resilience arises from the mission configuration. Developing such a method would pose challenges of optimizing the probability that data from each waypoint is collected and delivered to base.

Another area of related research into resilience of mission design would be to develop approaches supporting surveillance of a value-weighted map, where only a certain percentage of value is required to achieve mission success. This would require development of metric to determine the "value" of a waypoint, as well as formulation of an integer linear programming problem of maximizing the total value of waypoints visited. This would be an example of the *Multiple Traveling Thieves Problem*, another *NP*-hard integer linear programming problem [34].

This research also serves as a foundation for improved integration of advanced encryption techniques into UAS missions in the context of adversarial threats. UASs equipped with pretrained machine learning (ML) models could perform more sophisticated activities in the field, such as image classification, but carrying such a model exposes new attack surfaces [35]. A captured UAS could be exploited to reveal not only content but ML model parameters and other information regarding the UAS's specific mission. For surveillance applications whose goal is simply for UASs to collect data and return it to a secure base, collected data can be secured using *asymmetric encryption*, requiring a separately provided key to decrypt [36]. However, encrypting the activities of an ML model itself is more challenging. One approach would be to incorporate *Secure Multiparty Computation* (MPC) (e.g. [37], [38], [39], [40], [41]). MPC could be used to essentially "split" the ML model parameters into uninterpretable shares across a UAS swarm, requiring any combination of $K \leq N$ UASs (for some threshold $K$ depending on the particular MPC protocol) to be co-located and able to communicate to run the model. Implementing this type of mission is operationally challenging and requires updating the routing methods to collocate $K$ UASs at certain points in time and space but would greatly improve operational capabilities and UAS mission resilience to adversarial threats.

## V. CONCLUSION

In this study, a mathematically rigorous computational model was presented to support a UAS mission of collecting surveillance data from all parts of a contested or risky area and then returning this data to a home base. The solution to this problem requires a combination of several mathematical and statistical techniques. A path planning strategy ("Robust") was constructed that is robust to the possible disruption of UASs going offline due to hazard or threat. Using discrete probability theory, this strategy predicts the minimal number of UASs that should be deployed to ensure that the mission will be successful within a user-defined threshold probability of success. In comparison with the "Efficient" strategy, it was shown that the proposed method is more hardened against disruption. Further, results are favorable for a preliminary hybrid resilient approach ("Resilient") between the two strategies. This research may be useful to decision-makers who must survey risky areas with UASs or who have a limited number of UASs available for deployment. Above all, this study provides a case study that demonstrates that considering robustness as well as resilience in formulating the problem may entirely change the solving method, as well as the resulting optimal solution itself. Therefore, robustness and resilience should be formalized and incorporated in problem formulation at early stages.

## DATA AVAILABILITY

All data in this study was synthetically generated and is available upon request.

## COMPETING INTERESTS

The authors declare they have no competing interests.

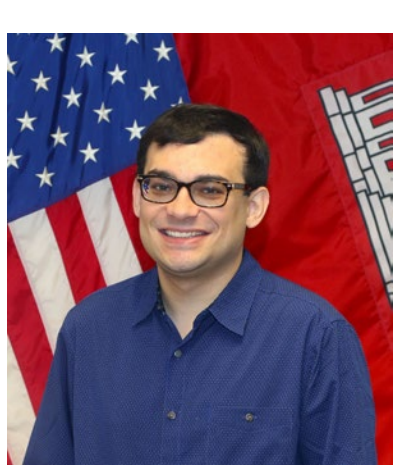

**Michael J. Gaiewski** received the B.A. in mathematics and economics in 2018 from Clark University in Worcester, MA USA. He then received the Ph. D. in mathematics from the University of Connecticut in Storrs, CT USA in 2024 studying finite element methods.

He currently works as a Research Scientist for Credere Associates LLC., providing services to the US Army Corps of Engineers Research & Development Center (ERDC). His current professional interests include numerical methods, computational mathematics, and mathematical algorithm development.

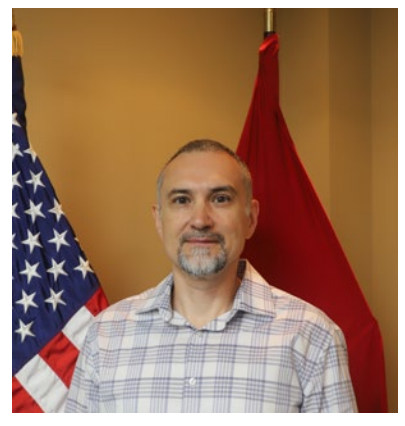

**Sergey N. Vecherin** received the M.S. in physics in 1998 from the Moscow State University, Russia. In 2007, he graduated from the New Mexico State University, Las Cruces, NM, USA with the Ph. D. degree in physics with the specialization in the acoustic inverse problems.

He is currently a Research Physicist with the US Army ERDC-CRREL. His current professional interests include forward and inverse problems in acoustic and seismic wave propagation, probabilistic analysis and modeling, and mathematical algorithm development for complex optimization problems.

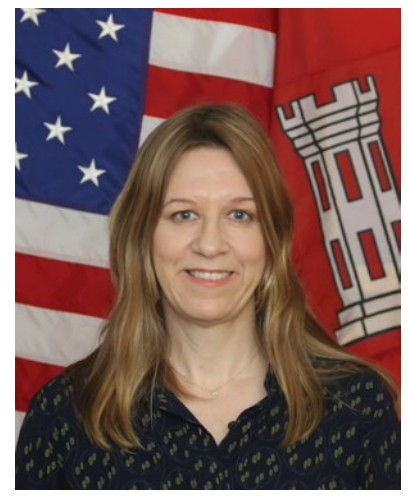

**Laurel P. Williams** received the M.S. degree in natural resources from the University of Vermont, Burlington, VT, USA, in 2006.

She has over two decades of experience in successful delivery of technical products and services within the public and private sectors, in areas including risk and decision analysis, natural resources, and enterprise information technology. She is a Program Manager with Credere Associates, LLC., providing project and program management and research services to the US Army Corps of Engineers Research & Development Center (ERDC). Her research interests lie in complexity, its expression within the built and natural environment, and the relationship of both to socio-ecological systems.

Ms. Williams is a certified PMI Project Management Professional and Certified Public Manager (CPM).

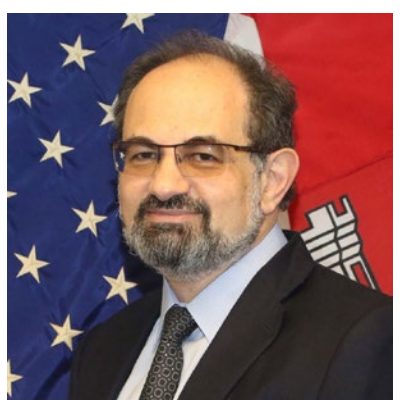

**Dr. Igor Linkov** is a Senior Executive with the US Army Engineer Research and Development Center (ERDC), and Adjunct Professor with University of Florida and Carnegie Mellon University. He develops methods and tools for measuring resilience in interconnected networks and applies these tools to the environment, critical infrastructure, transportation, energy and cyber systems, supply chains as well as command and control systems. He has published 30 books and over 500 peer reviewed papers and book chapters.